\documentclass[fleqn,12pt,twoside]{article}
\usepackage{espcrc1}

\usepackage{graphicx}
\usepackage[figuresright]{rotating}

\newcommand{\AmS}{{\protect\the\textfont2
  A\kern-.1667em\lower.5ex\hbox{M}\kern-.125emS}}

\title{The Muon Spectrometer of the ALICE experiment}

\author{Gin\'es MARTINEZ \address[suba]{SUBATECH (EMN-UN-IN2P3), 4 rue Alfred Kastler BP20722, 44307 Nantes Cedex 3} \\
for the ALICE Collaboration}
       
\begin{document}

\maketitle

\begin{abstract}
The main goal of the Muon Spectrometer of the ALICE experiment is the measurement 
of heavy quarks in pp, pA and AA collisions at LHC energies, via the muonic channel.
Physics motivations, the apparatus and its physics performances are presented in this talk.
\end{abstract}

\section{Physics motivations}

Heavy ion collisions at relativistic energies are a priviledge tool for creating very hot and dense matter in a laboratory.
In particular, lattice chromo-dynamics (lQCD) predicts a cross-over toward a new state of matter called Quark Gluon Plasma (QGP) 
at a temperature $\sim~170$ MeV for vanishing chemical potential $\mu_B$ \cite{Kars01}.
Heavy ion collisions allow to experimentally study the properties of this new state of matter.
This experimental program started in the mid 80s with fixed target heavy ion experiments at the AGS and 
SPS \cite{Satz02} 
and continued with the physics program developed at the RHIC collider (BNL) \cite{Hemm04}.
Heavy ion collisions at the future Large Hadron Collider (LHC) at CERN 
will open new experimental insights in the study of hadronic matter at high temperature.
The ALICE experiment will be the only experiment at LHC devoted to the heavy ion physics \cite{Safa04}, 
whereas the ATLAS and CMS experiments plan to develop a heavy-ion program \cite{Wysl04,Taka04} in parallel with their
main physics goal.

The LHC collider will provide proton and lead high luminosity beams at 7.0 TeV and 2.75A TeV momentum respectively.
At such ultra-relativistic energies new phenomena emerge, improving the experimental conditions for studying 
the hadronic matter in nucleus-nucleus central collisions: 
\begin{itemize}
\item \textbf{Initial conditions.} The initial conditions will be under control by the gluon saturation scenario.
At these energies the initial nucleus-nucleus interaction can be viewed as \emph{weak} 
interactions of a huge number of \emph{small x} gluons which will be freed in the beginning 
of the collision leading to a formation of 
a big gluonic ball \cite{Lerr01,Ianc03}.
In addition, most of these processes (like secondary interaction of minijets) leading to thermalization will be 
governed by hard processes ($\alpha_s<1$) 
which can be theoretically studied by perturbative chromo-dynamics (pQCD).  
\item \textbf{Equilibrated matter.} After equilibration of the initial gluonic ball, a hotter and longer-lived 
hadronic matter will be formed. The increase of the beam energy will favor the creation of vanishing baryonic potential  
hadronic matter with a temperature around 0.5-1 GeV,
well above the critical temperature predicted by lQCD.
\item \textbf{Observables.} Hadronic matter and collision dynamics will be probed with new observables 
which become only available with the increasing beam energy: 
event-by-event fluctuations, jet production, photon-jet correlations, open heavy flavor and Upsilon family resonance production.
\end{itemize}

\paragraph{Heavy Quark Production}
Charm quarks will be copiously produced in Pb+Pb collisions at $\sqrt{s}\sim$5.5A TeV: up 
to hundred of $c\bar{c}$ pairs per collision (and around 5 $b\bar{b}$ pairs per collision) \cite{Yell03}.
Production of heavy quarks will be dominated by prompt parton-parton scattering,
although new phenomena as, for instance, heavy quark production in secondary minijet interactions 
could noticeably contribute to the total production cross-section \cite{Mull92}.
These heavy quarks will be embedded in a matter mainly formed by gluons and light quarks (u,d,s) and could
behave like \emph{Brownian particles} \cite{Goss04}, therefore 
their transverse momentum (p$_T$) and rapidity distributions will probe the properties of the surrounding matter.
In addition, the study of the heavy quark bound states will allow to probe the medium via 
the Debye screening \cite{Datt04}, the gluon dissociation of quarkonia \cite{Gran04}, 
statistical recombination of heavy quarks \cite{Thew01} and/or statistical hadronization \cite{Brau00}.
Additionally, high p$_T$ heavy quarks will also probe the surrounding hadronic medium via heavy-quark matter 
interaction \cite{Zhan04}.

\section{Muon Spectrometer}

In the framework of the ALICE physics program \cite{Safa04}, 
the goal of the Muon spectrometer of ALICE is the study of open heavy flavor production and quarkonia production 
($J/\psi$, $\psi'$ and $\Upsilon$(1$S$), $\Upsilon$(2$S$) and $\Upsilon$(3$S$)) via the muonic channel.
For AA collisions, the dependence with the collision centrality and with the reaction plane  
(measured with the ALICE central barrel) will also be studied.

\begin{figure}
\begin{minipage}[t]{80mm}
\includegraphics[width=80mm,clip]{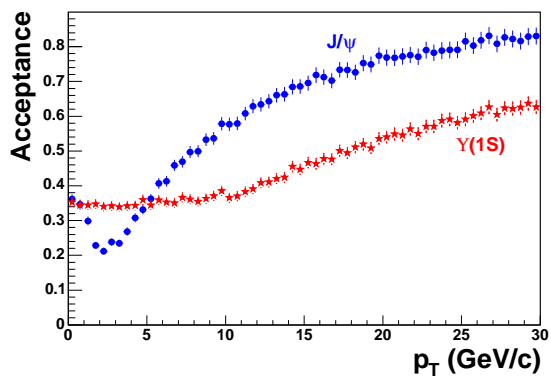}
\caption{
{\small Acceptance of the MUON spectrometer as a function of the transverse momentum for $J/\psi$ and 
$\Upsilon$(1S) in the rapidity range $-4.0<y<-2.5$, 
via their muon pair decay and with a muon low p$_T$ cut equal to 1 GeV/c.}
\label{fig:acceptance}}
\end{minipage}
\begin{minipage}[t]{80mm}
\includegraphics[width=80mm,clip]{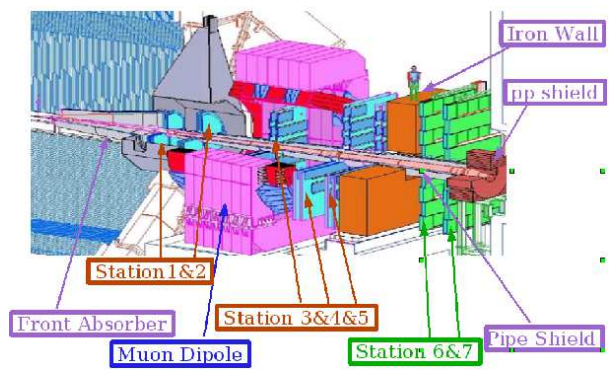}
\caption{
\small{Lay-out of the muon spectrometer. The tracking system consists of stations 1 to 5 and 
trigger system of stations 6 and 7.}}
\label{fig:muonspectrometer}
\end{minipage}
\end{figure}

The main experimental requirement is to measure the quarkonia production 
in central Pb+Pb collisions at LHC energies, down to very low p$_T$, since
low p$_T$ quarkonia will be sensitive to medium effects like heavy-quark potential screening.
Since muons are passively identified by the absorber technique,
a Lorentz boost is needed to be able to measure quarkonia at low p$_T$.
On the other hand, the muon spectrometer has to be as close as possible to the physics of the QGP 
which occurs in the mid-rapidity region.
As a compromise, the muon spectrometer allows for measuring muons and quarkonia production in an intermediate 
rapidity range $-4.0<y<-2.5$.
The acceptance plot in this rapidity range for $J/\psi$ and $\Upsilon$(1S) mesons decaying into muon pairs 
is presented in Fig.\ref{fig:acceptance}.
The muon spectrometer will be a unique apparatus at LHC to measure charmonia production 
at p$_T\sim0$
and will cover a rapidity range which completes the one measured by the ALICE central barrel, and by the CMS and ATLAS experiments.
The second main experimental requirement for the muon spectrometer is to be able to disentangle the different 
resonances of the $\Upsilon$ family. In particular the separation between the resonance $\Upsilon$(2S)
and $\Upsilon$(3S) ($\Delta$M $\sim$ 300 MeV/c$^2$) imposes to the spectrometer 
an  invariant mass resolution about 100 MeV/c$^2$ in the $\Upsilon$ mass region.

In order to reach these requirements, the muon spectrometer is located downstream (side C of point 2 at LHC) 
of the ALICE detector covering the angular range $171.^\circ<\theta<178.^\circ$,
consisting of 3 absorbers, a muon magnet, a trigger system and a tracking system.
The lay-out of the muon spectrometer is presented on Fig. \ref{fig:muonspectrometer} \cite{Muon99}.

\subsection{Absorbers}
Absorbers reduce the initial flux of primary hadrons from nucleus-nucleus collisions by a factor $\sim$100, 
and they protect the detectors from low energy particles created in secondary interactions 
(mainly low energy electrons).
The front absorber is the most critical component and it has been designed for minimizing 
the invariant mass resolution deterioration of the spectrometer due to straggling and multi-scattering.
This imposes an upper limit of the amount of material ($\lambda_I\sim 10$), 
and requires that components with low Z are located close to the interaction point (IP), whereas high Z components 
are placed close to the spectrometer.
Moreover, muons from hadronic weak decay are optimally suppressed by placing the front absorber as close as possible 
to the interaction point. The distance from the IP is, however, limited to 90 cm since physics performance of 
the ALICE central barrel should not be deteriorated by the presence of the absorber.
The front absorber is the main contributor to the invariant mass resolution of the spectrometer
in the $\Upsilon$ region, with a quadratic contribution equal to about 80 MeV/c$^2$.
The absorber around the beam pipe is crucial to reduce the low energy background in the tracking and trigger chambers 
due to secondary interactions of beam particles in the pipe.
Finally, an iron wall 120 cm thick located between the tracking stations (stations 1-5 in fig.\ref{fig:muonspectrometer}) 
and the trigger stations  (stations 6 and 7 in fig.\ref{fig:muonspectrometer})  allows for reducing 
the low energy background in the trigger chambers which are less constrained by straggling and multi-scattering.
At present (Sept 2004), the absorbers are in the construction phase at CERN.

\subsection{Muon magnet}
Muon momenta are determined by muon tracking in a magnetic field generated by a warm dipole of 820 tons, 
nominal field of 0.7 T and a field integral along beam axis $\int |B| dz \sim 3$ Tm.
The magnetic field is directed in the horizontal plane perpendicular to the beam direction 
(x axis) defining a bending plane (zy plane) and a non bending plane (xz plane).
The muon magnet has been assembled and is being tested at CERN, final assembly is foreseen in 2005. 

\subsection{Trigger system}
Muon detection based on absorption technique allows for a very efficient triggering on high p$_T$ muons.
This is crucial in order to take advantage of the full luminosity of the heavy ion beams at LHC, taking into account 
that muon acquisition system is limited to an event rate of 1kHz.
The trigger system (stations 6 and 7 in fig.\ref{fig:muonspectrometer}) consists of 4 planes of 18 Resistive Plate Chambers (RPC) each, 
located between 16 m and 17 m downstream (just behind the iron wall) and 
operating in the streamer mode with a gas mixture of Ar, CH$_2$F$_4$, 
C$_4$H$_{10}$ and SF$_6$ \cite{Arna00} \footnote{For the pp runs, an avalanche operating mode is being investigated due to trigger chamber ageing problems}.
Signals in individual strips of the RPC are treated by a dual threshold discriminator (ADULT) \cite{Arna01}.
Information from the 4 trigger detection planes are locally processed by the local hardware cards, 
determining roughly the transverse momentum of the muon track.   
Regional and global hardware cards collect the full information from local cards and determine the trigger condition 
of the event in less than 700 ns. 
Different trigger types are possible: 
low p$_T$ (above 1 GeV/c for $J/\psi$ studies), high p$_T$ (above 2 GeV/c for $\Upsilon$ studies),
unlike or like sign muon pairs and
single muons.
The muon trigger delivers information to the central trigger processor for the generation of the ALICE level 0 trigger.

\subsection{Tracking system}
\begin{figure}
\begin{minipage}{65mm}
\includegraphics[width=6.5cm,clip]{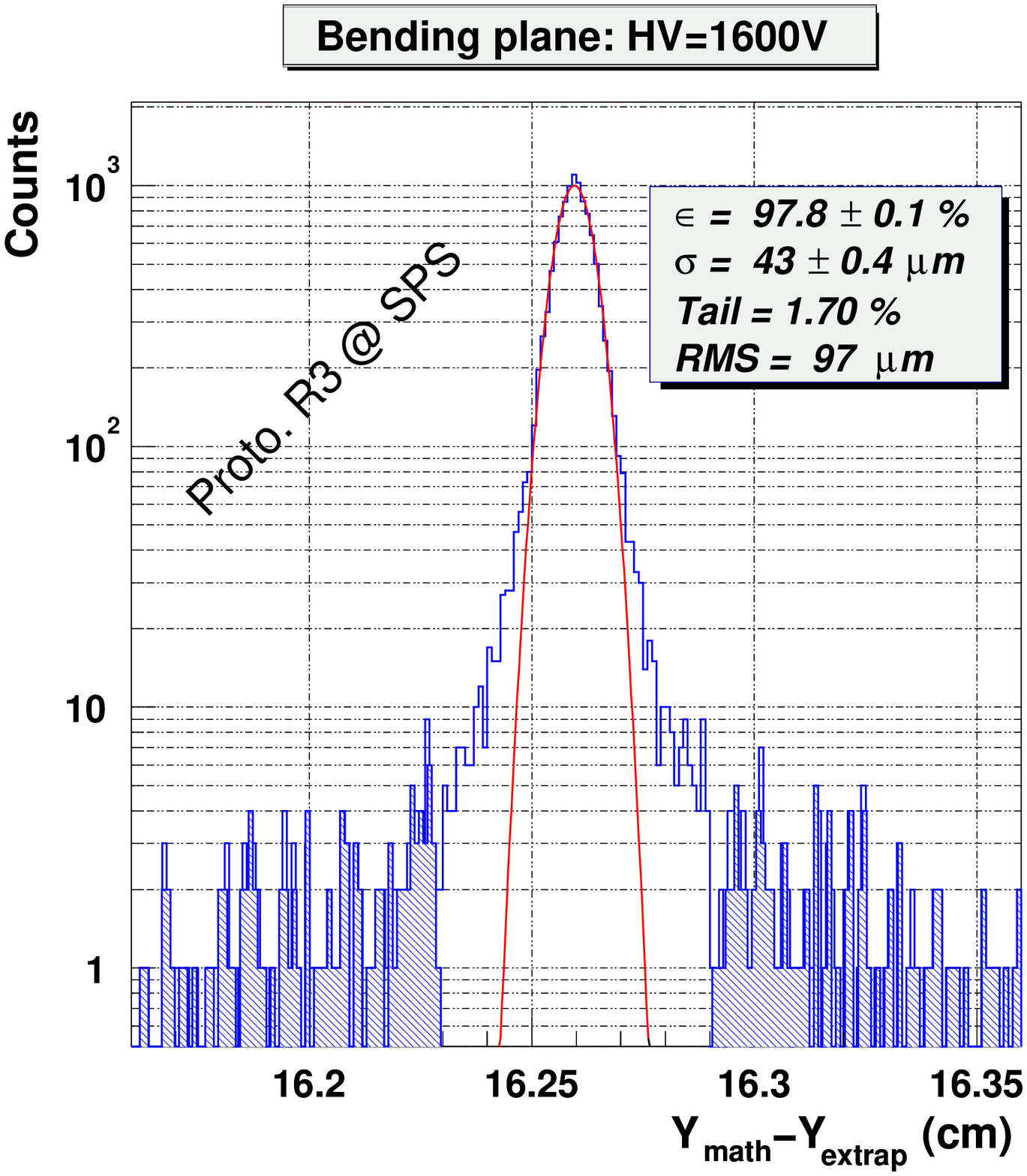}
\caption{\small{Spatial resolution in the bending plane of the CPC under muon beam test at 100 GeV/c in SPS.}
\label{fig:resolution}}
\end{minipage}
\begin{minipage}{90mm}
\includegraphics[width=9cm,clip]{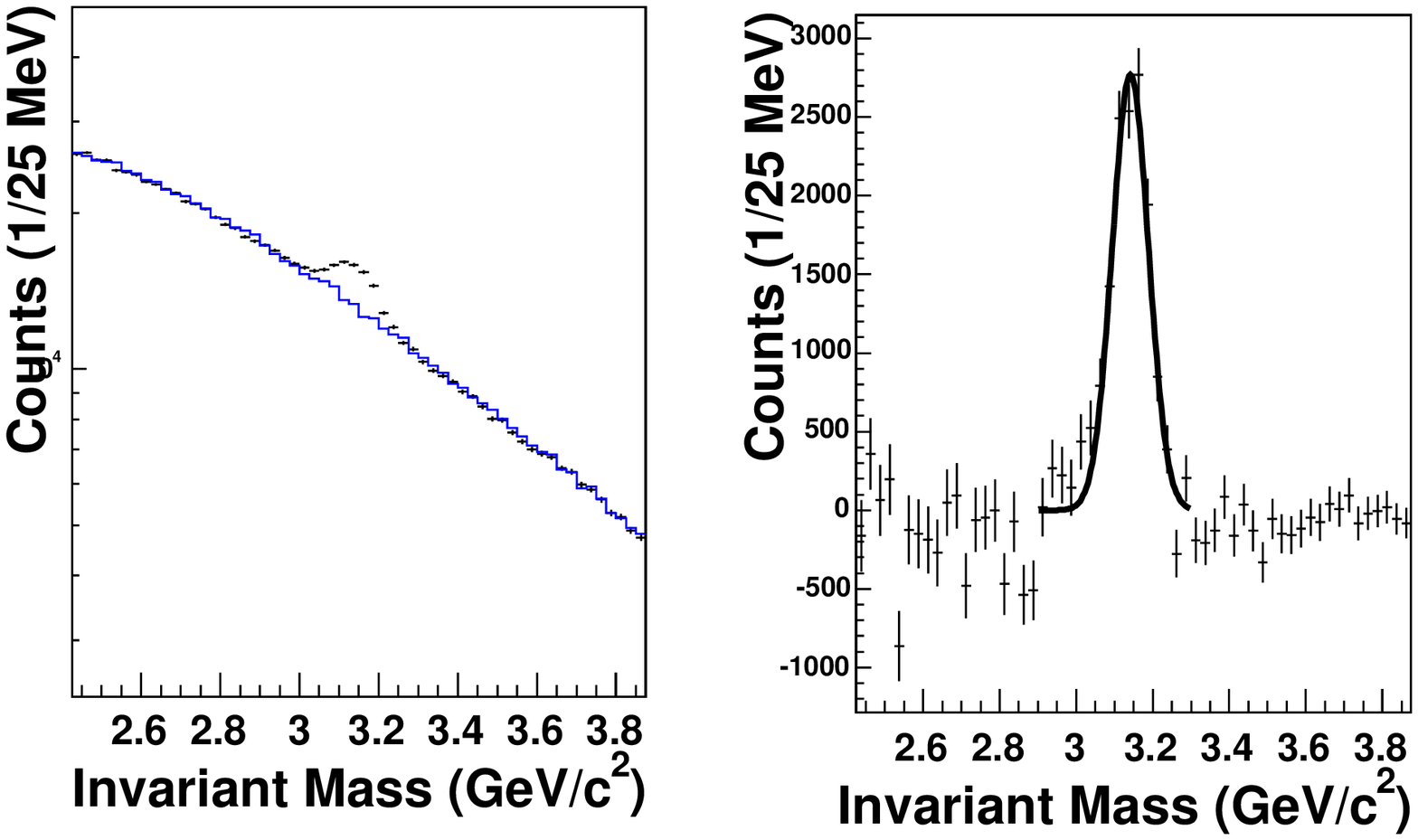}
\caption{\small{Invariant mass distribution of muon unlike pairs in the Muon Spectrometer for Pb+Pb collisions at 
LHC energies in the most central collisions 0-10\%.
The simulated statistics correspond to about 7\% of the full statistics of 1 year PbPb run ($10^6$ s) in LHC 
at nominal luminosity.}}
\label{fig:JpsiCentral}
\end{minipage}
\end{figure}

Muon transverse momenta in the bending plane ($|$p$_{zy}|$) 
are determined by tracking muons along the magnetic field. 
A momentum resolution about 1\% is needed to achieve the required resolution in the $\Upsilon$ 
invariant mass region ($\sim$ 100 MeV/c$^2$). This imposes a spatial resolution of the tracking system 
in the bending plane better than 100 $\mu$m  and a tracking system with a reduced radiation thickness.
The tracking system is made of 5 stations with 2 detection planes consisting of 5mm drift 
multi-wire proportional chambers with bi-cathode pad read-out (cathode pad chambers, CPC).
Thickness of each chamber is below 3\% radiation length. 
The first 2 stations are placed in front of the muon magnet at a distance of $\sim$5.4 m and $\sim$6.8 m respectively 
from the IP. 
They consist of 4 detection planes made of  4 CPC each with quadrant design.
Stations 3, 4 and 5 are placed at a distance of $\sim$9.7 m (inside the muon magnet),   
$\sim$12.65 m and $\sim$14.25 m from the IP.
A modular design has been chosen for these stations, consisting of rectangular CPC called \emph{slats} 
(36, 52 and 52 slats for stations 3, 4 and 5 respectively).
Different pad densities are present in the CPC depending on the station and pseudo-rapidity positions, 
ranging from 5$\times$6 mm$^2$ for pads closest to the beam-pipe in station 1, to 5$\times$100 mm$^2$ 
for stations 3,4,5 at low pseudo-rapidities.
An initial direction of the muon is determined from the track parameters of the muon and from the position of the IP, taking into 
account the multi-scattering in the absorber. 
In order to keep the invariant mass resolution in the $\Upsilon$ region below 
100 MeV/c$^2$, a spatial resolution of 1 mm is required in the non-bending plane (y direction) 
as well as a position resolution of the IP along the beam axis below 1 cm.
A IP position will be provided by the pixel internal tracking system (ITS) of ALICE \cite{ITS}.
CPC chambers have been tested in beam at PS and at SPS (CERN) and the spatial resolution in the bending-plane 
is better than the required value (see Fig.\ref{fig:resolution}), with an efficiency of 98\%.
The muon tracking system consists of  10$^6$ channels which are stored in 20 subevents (4 per station or 2 per detection plane).
Signal induced in the CPC pads is read-out by the FEE MANU card, based on MANAS pre-amplifiers and a MARC ADCs.
Data is then serialized and concentrated by the CROCUS and data concentrator cards (1 CROCUS card per subevent).

\section{Physics studies}
The Muon physics program is focused on the measurement of heavy flavor production.
Many studies have been undertaken mainly in the framework of the Technical Design Report \cite{Muon99} and 
the Physics Performace Report of ALICE \cite{PPR} and other studies are still in progress.  
\subsection{Quarkonia measurements}
Invariant mass analysis of muon  pairs will provide a direct measurement of the quarkonia production.
$J/\psi$, $\Upsilon$(1S), $\Upsilon$(2S), $\Upsilon$(3S) will be measured in pp, pA and AA collisions from the most central 
to the most peripheral collisions. Quarkonia measurement in central Pb+Pb collisions will be very challenging due to the huge low-energy 
background and the large muon combinatorial background. In the case of $J/\psi$, a ratio signal to background (S/B) 
in the order of 0.1 is expected in central (0-10\%) Pb+Pb collisions at 5.5A TeV (see Fig.\ref{fig:JpsiCentral}).
During the first Pb+Pb run with nominal luminosity at LHC (this corresponds to a period of 10$^6$ s and  a luminosity of $5\cdot 10^{26}$ cm$^{-2}$ s$^{-1}$) 
we expect to measure about half a million of $J/\psi$ 
and 6400 $\Upsilon$(1S), assuming a scaling on the number of collisions in AA collisions. 
Those numbers would fluctuate depending on the
physics of the heavy quark production and quarkonia production in heavy ion collisions: 
from total suppression due to color screening, to enhancement due to recombination.
The muon trigger system will allow for taking advantage of the full luminosity of the heavy ion beams at LHC. 
Trigger rates for Pb+Pb collisions will be around 500 s$^{-1}$.
Measurement of the ratio of  $\Upsilon$(2S)/$\Upsilon$(1S)  will provide a very powerful experimental observable to constraint 
the different models on quarkonia suppression in the QGP \cite{Gun97}.
The measurement of $J/\psi$ elliptic flow will also be possible.
Finally, the study of very high p$_T$ $J/\psi$ will represent a very promising hard probe.

\subsection{Open heavy flavor measurements}
The measurement of open heavy flavors will also be a priority of our physics program.
Being very interesting in its own, it will allow for a normalization of the quarkonia production rates.
Different alternative analysis will be applied, which exploit the muon production from open charm and beauty mesons via their semi-leptonic
decay.
\begin{itemize}
\item The measurement of muon p$_T$ distribution will provide the first measurement of heavy quark 
production at high rapidities at LHC. 
For p$_T$ larger than 5-8 GeV/c muon production will be dominated by the semi-muonic decay of D and B mesons.

\item Like and unlike sign muon pairs originating from the same hard scattering or same heavy quark, will present a residual 
correlation in the low (1-3 GeV/c$^2$) and high (4-8 GeV/c$^2$) invariant mass ($M_{inv}$) regions. 
Correlated unlike-sign muon pairs in high $M_{inv}$ region will be mainly produced
by semi-muonic decays of $D-\bar{D}$ and  $B-\bar{B}$ mesons from the same hard scattering,
whereas the low $M_{inv}$ region will be populated by $B-D$ or $\bar{B}-\bar{D}$ semi-muonic decays from the same heavy quark fragmentation.
Like-sign muon pairs will be dominated by $B-\bar{D}$ and, more exotic mechanism, like 
$B^0-\bar{B}^0$ oscillations.

\item Multi-correlation of muons (3 or 4) mainly streaming  from the beauty production, 
correlations with electrons and/or kaons in the central ALICE barrel produced by heavy quark production in an intermediate 
rapidity region, etc ...  will present alternative experimental methods to measure the heavy quark production.
\end{itemize}

\section{Conclusions and Perspectives}
In summary, the muon spectrometer supports an ambitious physics program in the ALICE experiment, focused on heavy quark physics 
for studying QGP at the LHC collider.
The muon spectrometer has entered in its construction phase and will be ready in 2007 at the interaction point 2 of LHC 
for the first data taking at LHC.

\section{Acknowledgments}
I would like to thank my colleagues of ALICE collaboration for fruitful discussions.
In particular, I would like to thank Philippe Crochet, Eric Dumonteil, Christian Finck, Fabien Guerin, 
Smbat Grigoryan, Andreas Morsch, Karel Safarik and Alain Tournaire for helping me in the preparation of this talk. 
I would like to thank Philippe Crochet, Christian Finck, Hans-Ake Gustafsson and Jacques Martino for carefully reading this manuscript.

\end{document}